\documentclass[aps,prl,twocolumn]{revtex4}
\makeatletter

\newcommand{\Rmnum}[1]{\expandafter\@slowromancap\romannumeral #1@}
\makeatother
\usepackage{graphicx}
\usepackage{dcolumn}
\usepackage{bm}
\usepackage{CJK}
\usepackage{color}
\usepackage{amsfonts}
\usepackage{psfrag}
\usepackage{wrapfig}
\usepackage{subfigure}
\usepackage{makeidx}
\usepackage{multirow}
\usepackage{epsf}
\usepackage[colorlinks,linkcolor=red,anchorcolor=red,citecolor=blue,urlcolor=blue]{hyperref}
\usepackage{amsmath}
\usepackage{cases}
\usepackage{bm}

\begin{document}
\begin{CJK}{GBK}{song}
\title{Fundamental and second-order super-regular breathers in vector fields}

\author{Chong Liu$^{1,2,3,4}$}\email{chongliu@nwu.edu.cn}
\author{Shao-Chun Chen$^{1}$}
\author{Nail Akhmediev$^{2,5}$}
\address{$^1$School of Physics, Northwest University, Xi'an 710127, China}
\address{$^2$ Department of Fundamental and Theoretical Physics, Research School of Physics, The Australian National University, Canberra, ACT 2600, Australia}
\address{$^3$Shaanxi Key Laboratory for Theoretical Physics Frontiers, Xi'an 710127, China}
\address{$^4$NSFC-SPTP Peng Huanwu Center for Fundamental Theory, Xi'an 710127, China}
\address{$^5$Arts $\&$ Sciences Division, Texas A$\&$M University at Qatar, Doha, Qatar}

\begin{abstract}
We developed an exact theory of the super-regular (SR) breathers of Manakov equations.
We have shown that the vector SR breathers do exist both in the cases of focusing and defocusing Manakov systems. The theory is based on the eigenvalue analysis and on finding the exact links between the SR breathers and modulation instability. We have shown that in the focusing case the localised periodic initial modulation of the plane wave may excite both a single SR breather and the second-order SR breathers involving four fundamental breathers.
\end{abstract}

\maketitle

\textit{Introduction}{.---} Understanding the formation of oscillating localised structures known as `breathers' is a fundamental problem in a wide variety of conservative and dissipative systems \cite{Book97,Book05,Book08,DB1,DB2,Lattice,PT}.
Breathers are known in optics \cite{Dudley1}, hydrodynamics \cite{Dudley2}, Bose-Einstein condensates \cite{Lattice}, micromechanical arrays \cite{Sato}, and in the cavity optomechanics \cite{Wu}.
They provide a basis for more complicated formations---nonlinear superpositions of breathers that appear in many nonlinear phenomena of physical importance such as rogue wave events \cite{RW2009,Extreme09,Excite09}, breather molecules \cite{BS}, chess-board-like patterns \cite{CBL}, breather turbulence \cite{Turbulence}, higher-order modulation instability (MI) \cite{JETP88,Exp2011-PRL}, and the MI where small periodic modulation is additionally localised in transverse direction.
The latter leads to the excitation of super-regular (SR) breathers \cite{SR1}.

As it was shown in \cite{SR1,SR2,SR3}, the
SR breathers are the higher-order exact solutions of the scalar nonlinear Schr\"odinger equation (NLSE)  that consist of two fundamental breathers propagating at small angle to each other.
Later, the ideas of SR-breather theory have been successfully applied to the NLSE with higher-order effects \cite{CL0}, complex modified KdV equation \cite{CL1}, equations modelling the resonant erbium-doped fiber \cite{CL3}, self-induced transparency \cite{CL4}, and the derivative NLSE \cite{CL5}.
One of the important results of these studies is that the MI growth rate related to the excitation of the SR breathers is defined by the absolute difference of the group velocities of the two breathers \cite{CL2}.

SR breathers modelled by the scalar NLSE have been observed both in experiments in fiber optics and on water surface \cite{SR3}. In each medium, the initial conditions required to excite the SR breathers have been carefully modelled by the exact solutions. Further studies have shown that a larger variety of initial conditions also lead to their excitation \cite{AG,MC,ST}. So far, these studies are limited to the NLSE-type integrable systems with an associated $2\times2$ Lax pair. Extending this knowledge to the more complex systems consisting of two coupled wave fields still remains a challenge.
One of the practically important cases involves Manakov equations \cite{MM}. These equations play a pivotal role in modelling variety of nonlinear wave phenomena in Bose-Einstein condensates \cite{BEC}, in optics \cite{OF,MS1996,VRW2016}, in hydrodynamics \cite{F}, and, perhaps, with some reservations, in finances \cite{Yan}.

Despite being vigorously investigated, higher-order nonlinear excitations in coupled wave systems still require more efforts. Several preliminary studies in this direction have been made \cite{VSR1, VSR2}.
The complications arise because of the increased number of spectral parameters in the vector breather solutions. Also, the practically important task of  their excitation from  weak modulations remains to be addressed.

\textit{Breather solutions of Manakov equations}{.---} We start with the Manakov equations in dimensionless form, which are given by
\begin{eqnarray}\label{eq1}
\begin{split}
i\frac{\partial\bm\psi^{(1)}}{\partial t}+\frac{1}{2}\frac{\partial^2\bm\psi^{(1)}}{\partial x^2}+\sigma \left(\bm|\bm\psi^{(1)}\bm|^2+\bm|\bm\psi^{(2)}\bm|^2 \right)\bm\psi^{(1)}&=0,\\
i\frac{\partial\bm\psi^{(2)}}{\partial t}+\frac{1}{2}\frac{\partial^2\bm\psi^{(2)}}{\partial x^2}+\sigma\left(\bm|\bm\psi^{(1)}\bm|^2+\bm|\bm\psi^{(2)}\bm|^2\right)\bm\psi^{(2)}&=0,
\end{split}
\end{eqnarray}
where $\bm\psi^{(1)}(t,x)$, $\bm\psi^{(2)}(t,x)$ are the two nonlinearly coupled components of the vector wave field.
Parameter $\sigma(=\pm1)$ defines the strength of the nonlinear terms in Eqs. (\ref{eq1}).
It corresponds (in optics) to the self-focusing regime when $\sigma=+1$ and self-defocusing regime when $\sigma=-1$.
The physical meaning of independent variables $x$ and $t$ depends on a particular physical problem of interest. In optics, $t$ is commonly a normalised distance along the fibre while $x$ is the normalised time in a frame moving with the group velocity. In the case of Bose-Einstein condensates, $t$ is time while $x$ is the spatial coordinate.

The fundamental vector breather solution can be constructed by using the Darboux
transformation method \cite{DT1991}. Its general form reads
\begin{eqnarray}
\bm\psi^{(j)}(t,x)=\bm\psi_{0}^{(j)}(t,x)\left[1+(\bm\lambda^*-\bm\lambda)\bm\psi_a^{(j)}(t,x)\right],\label{eqb}
\end{eqnarray}
where $\bm\psi_{0}^{(j)}=a_j \exp \left\{ \left\{i\beta_jx + i\left[\sigma( a_1^2+ a_2^2)- \frac{1}{2}\beta_j^2 \right]t\right\} \right\}$ is the vector plane wave of Eqs. (\ref{eq1})
with $a_j$, $\beta_j$ being the amplitudes and the wavenumbers, respectively.
From a physical perspective, the relative wavenumber is important since it cannot be eliminated through Galilean transformation.
Without losing generality, we set $\beta_1=-\beta_2=\beta$.
Parameter $\bm\lambda$ stands for the spectral parameter of the Lax pair of Eqs. (\ref{eq1}) and the asterisk denotes the complex conjugation.
It is given by $\bm\lambda=\bm\chi-\sigma\sum_{j=1}^2a_j^2/(\bm\chi+\beta_j)$
with $\bm\chi$ being the corresponding eigenvalue. The latter obeys the relation
\begin{eqnarray}
1+\sigma\sum_{j=1}^2\frac{a_j^2}{(\bm\chi+\beta_j)(\tilde{\bm\chi}+\beta_j)}=0,\label{eqchirelation}
\end{eqnarray}
where $\tilde{\bm\chi}=\bm\chi+2\alpha$ with $\alpha$ being an arbitrary complex number.
Here, we fix
\begin{equation}
\alpha=\bm\omega/2+i\bm{\gamma}.
\end{equation}
Parameters $\bm\omega$ and $\bm{\gamma}$ determine the period and the width of the breather.
Moreover, $\bm\psi_a^{(j)}$ are given by
\begin{eqnarray}
\bm\psi_a^{(j)}=
\frac{\mathcal{B}^{(j)}(\bm\chi)\left(e^{\bm{\Gamma}}+e^{-i\bm{\Lambda}}\right)+\mathcal{B}^{(j)}(\tilde{\bm\chi})\left(e^{-\bm{\Gamma}}
+e^{i\bm{\Lambda}}\right)}
{\bm{\varepsilon}(\bm\chi)e^{\bm{\Gamma}}+\bm{\varepsilon}(\tilde{\bm\chi})e^{-\bm{\Gamma}}
+\mathcal{D}e^{i\bm{\Lambda}}+\mathcal{D}^*e^{-i\bm{\Lambda}}},\label{eqpa}
\end{eqnarray}
where the arguments $\bm\Gamma$ and $\bm\Lambda$ are
\begin{eqnarray}
\bm\Gamma=2\bm{\gamma}\{\bm x-V_g(\bm\chi)\bm t\},~~~\bm\Lambda=\bm\omega \{\bm x-V_p(\bm\chi)\bm t\}-\theta_1.
\end{eqnarray}
Here $\bm{x}=x-x_{01}$, $\bm{t}=t-t_{01}$ are shifted spatial and time variables respectively with $x_{01}$ and $t_{01}$ being responsible for the spatial and temporal position of the breather. Real parameter $\theta_1$ is an arbitrary phase.
Moreover, $V_g$ and $V_p$ denote the group and phase velocities, respectively, which are given by
\begin{eqnarray}
V_g(\bm\chi)&=&-\frac{\bm\omega}{2\bm{\gamma}}\bm\chi_i - \bm\omega - \bm\chi_r,\label{vg0}\\
V_p(\bm\chi)&=&\frac{2\bm\gamma}{\bm{\omega}}(\bm\chi_i+\bm{\gamma})-\frac{1}{2}\bm\omega-\bm\chi_r.\label{vp0}
\end{eqnarray}
Subscripts $r$ and $i$ denote the real and imaginary parts of the eigenvalue, respectively. The coefficients in (\ref{eqpa}) are:
$\mathcal{B}^{(j)}(\bm\chi)=1/(\bm\chi+\beta_j)$, $\bm{\varepsilon}(\bm\chi)=1+\sigma\sum_{j=1}^2\bm{|}a_j/(\beta_j+\bm\chi)\bm{|}^2$, and $\mathcal{D}=1+\sigma\sum_{j=1}^2a_j^2/[(\beta_j+\bm\chi^*)(\beta_j+\tilde{\bm\chi})]$.

The family of solutions (\ref{eqb}) depends on the plane-wave parameters ($a_j$, $\beta_j$), the real constants ($\bm{\gamma}$, $\bm\omega$), and the constant phase $\theta_1$.
It describes the growth-decay cycle of periodic structure on top of the plane wave. This structure
has the period $2\pi/\bm\omega$ and the width of the envelope $1/(2\bm{\gamma})$ in $x$.
It is propagating with the group velocity $V_g(\bm\chi)$ and the phase velocity $V_p(\bm\chi)$.

The family of solutions (\ref{eqb}) contains several subsets  \cite{VAB2021,VAB2022,VAB-Df2022,VRW-2022,VKMS-f2022}. They are vector generalizations of Kuznetsov-Ma solitons \cite{KM}, Akhmediev breathers (ABs) \cite{AB}, and Peregrine rogue waves \cite{Peregrine}.
Among them, we can mention the vector ABs, which exist when $\bm{\gamma}=0$, implying $\bm\Gamma=-\bm\omega \bm\chi_i t$.
They are localized in $t$ but periodic in space $x$ with period $2\pi/\bm\omega$.
The AB is known as the nonlinear stage of the MI developed from purely periodic modulation \cite{AB}.
The MI growth rate described by the AB solution is given by
\begin{eqnarray}
\label{eq-gab}
\mathcal{G}=|2\bm{\gamma}V_g(\bm\chi)|=\bm|\bm\omega \bm\chi_{_{AB}}\bm|,~~\bm\chi_{_{AB}}\equiv\bm\chi_i\bm|_{\bm{\gamma}=0}.
\end{eqnarray}

When $\bm{\gamma}\rightarrow0$, Eqs. (\ref{eqb}) describe the vector quasi-ABs with finite envelope width $1/(2\bm{\gamma})$ in $x$. The period in $x$ is still $2\pi/\bm\omega$.
The nonlinear superposition of a pair of quasi-ABs each with the period $2\pi/\bm\omega$ and the width of the envelope $1/(2\bm{\gamma})$ in $x$ is the SR breather. It describes the plane wave instability with the finite width of modulation.
The collision centre of the two quasi-ABs has the waveform of the  initial localised modulation.
The frequency of this localised modulation is within the MI band just like in the case of the AB.

Before entering the details of the analysis of the eigenvalue, let us recall the nature of the MI described by the SR breathers.
For a given physical system, both the SR breathers and the ABs are the nonlinear stages of the linear MI approximation.
They have the same growth rate $\mathcal{G}=\bm|\bm\omega \bm\chi_{_{AB}}\bm|$ and the same MI band that are found in the linear stability analysis that uses the  modulation frequency $\bm\omega$.
However, the physical meaning of the growth rate is more elaborate.
In our previous work \cite{CL2} involving $2\times2$ Lax matrixes, we have revealed
the exact link between the SR breathers and the MI. Namely, the growth rate defined by the SR breathers $\mathcal{G}$
is equal to the absolute difference of the group velocities of the paired quasi-ABs $\Delta V_g$. In other words, $\mathcal{G}=\bm\gamma\Delta V_g$.
Below, we demonstrate that an effective way to build the comprehensive concept of vector SR breathers
is to integrate such link and the eigenvalue analysis presented here.

\textit{Eigenvalue analysis of the vector SR breathers}{.---}From Eq. (\ref{eqchirelation}), we have explicit expressions for the eigenvalues $\bm\chi$:
\begin{eqnarray}\label{eqchi-SR}
\begin{split}
\bm\chi_1=(\bm\mu-\sqrt{\bm\nu})^{1/2}-\alpha,~\bm\chi_2=-(\bm\mu-\sqrt{\bm\nu})^{1/2}-\alpha,~\\
\bm\chi_3=(\bm\mu+\sqrt{\bm\nu})^{1/2}-\alpha,~\bm\chi_4=-(\bm\mu+\sqrt{\bm\nu})^{1/2}-\alpha,~
\end{split}
\end{eqnarray}
where
\begin{eqnarray}\label{eqmu}
\bm\mu=\alpha^2+\beta^2-\sigma a^2,~~\bm\nu=4(\alpha\beta)^2-4\sigma(a\beta)^2+a^4.
\end{eqnarray}
Expanding Eqs. (\ref{eqchi-SR}) using the small parameter $\bm{\gamma}$ ($\bm{\gamma}^2\ll1$), and separating the real and imaginary parts, in the first order, we have
\begin{eqnarray}\label{eqchi-SR0}
\begin{split}
\bm\chi_{1i}=\mathcal{O}(\bm\gamma^2)+\bm\chi^{(-)}_{_{AB}}-\bm\gamma,~\bm\chi_{1r}=\mathcal{O}(\bm\gamma^2)+\bm\rho^{(-)}-\bm\omega,\\
\bm\chi_{2i}=\mathcal{O}(\bm\gamma^2)-\bm\chi^{(-)}_{_{AB}}-\bm\gamma,~\bm\chi_{2r}=\mathcal{O}(\bm\gamma^2)-\bm\rho^{(-)}-\bm\omega,\\
\bm\chi_{3i}=\mathcal{O}(\bm\gamma^2)+\bm\chi^{(+)}_{_{AB}}-\bm\gamma,~\bm\chi_{3r}=\mathcal{O}(\bm\gamma^2)+\bm\rho^{(+)}-\bm\omega,\\
\bm\chi_{4i}=\mathcal{O}(\bm\gamma^2)-\bm\chi^{(+)}_{_{AB}}-\bm\gamma,~\bm\chi_{4r}=\mathcal{O}(\bm\gamma^2)-\bm\rho^{(+)}-\bm\omega,
\end{split}
\end{eqnarray}
where ($\bm\chi^{(-)}_{_{AB}}$, $\bm\chi^{(+)}_{_{AB}}$), ($\bm\rho^{(-)}$, $\bm\rho^{(+)}$) are the imaginary and real parts of the complex parameters when $\bm{\gamma}=0$. Namely,
\begin{eqnarray}\label{eqchi-AB}
\begin{split}
&i\bm\chi^{(-)}_{_{AB}}+\bm\rho^{(-)}=(\bm\mu-\sqrt{\bm\nu})^{1/2}\bm|_{\bm{\gamma}=0},\\
&i\bm\chi^{(+)}_{_{AB}}+\bm\rho^{(+)}=(\bm\mu+\sqrt{\bm\nu})^{1/2}\bm|_{\bm{\gamma}=0}.
\end{split}
\end{eqnarray}
Here, ($\bm\chi^{(-)}_{_{AB}}$, $\bm\chi^{(+)}_{_{AB}}$) determine the MI growth rate of the ABs, Eq. (\ref{eq-gab}), and their validity would determine further the validity of the eigenvalues of the vector SR breathers (\ref{eqchi-SR0}).

The nonlinear superposition of one pair of quasi-ABs with two different eigenvalues given by Eqs. (\ref{eqchi-SR0}) produces potentially a vector SR breather.
However, not every kind of combination of the eigenvalues leads to the SR-breather generation.
Below, we clarify this point analytically and numerically for both the defocusing and the focusing cases.

\begin{figure}[htb]
\centering
\includegraphics[width=85mm]{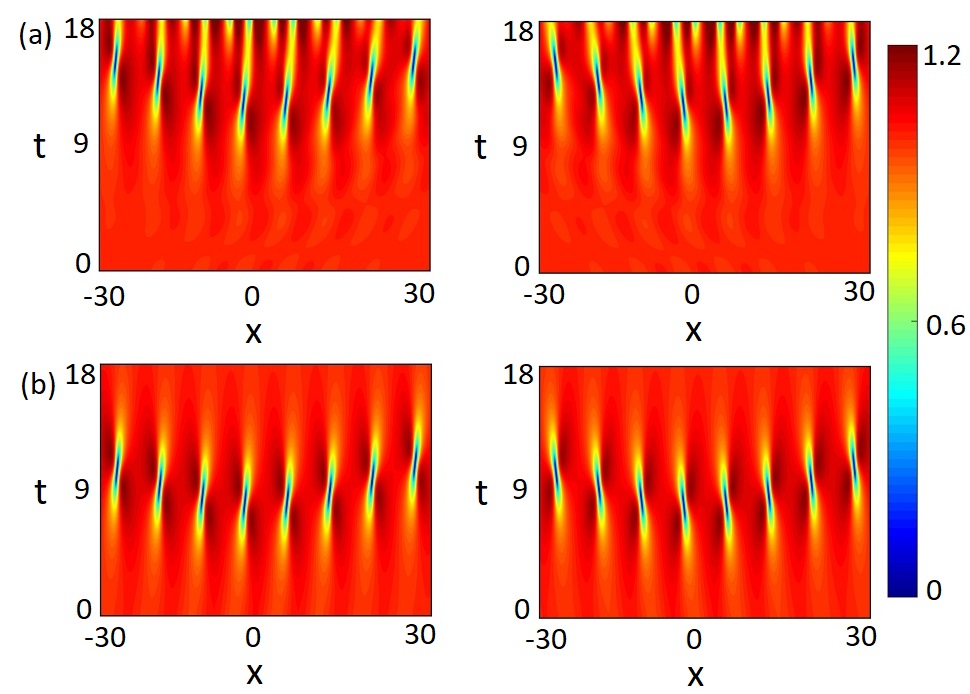}
\caption{Evolution of the amplitude profiles of the two components $|\bm\psi^{(j)}|$ of the vector SR breathers in the defocusing regime ($\sigma=-1$) obtained from (a) the numerical simulations started with the initial condition (\ref{eqic}), and (b) the exact solution $|\bm\psi^{(j)}(\bm\chi_{1};\bm\chi_{2})|$. Parameters are: $a_1=a_2=1$, $\beta=1$, $\bm\gamma=0.059$, $\bm\omega=0.8$, $\delta x=\{-4.8177,-2.6905\}$, $\delta t=\{0.6511,-0.0300\}$, $\theta_1=0$, $\theta_2=\pi$, and $x_{_W}=15$, $\phi^{(j)}=0$, $\epsilon=0.01$.}\label{f-defocusing}
\end{figure}

\textit{Vector SR breathers and their excitation in the defocusing regime}{.---} Let us first consider the defocusing case with $\sigma=-1$. Clearly, when $\sigma=-1$, we have $\bm\mu|_{\bm{\gamma}=0}>0$, $\bm\nu|_{\bm{\gamma}=0}>0$ from Eq. (\ref{eqmu}). This yields, from Eq. (\ref{eqchi-AB}), $\bm\chi^{(+)}_{_{AB}}=0$, but $\bm\chi^{(-)}_{_{AB}}\neq0$ (when $(\bm\mu-\sqrt{\bm\nu})\bm|_{\bm{\gamma}=0}<0$).
Thus, only eigenvalues $\bm\chi_1$, $\bm\chi_2$ are valid for the quasi-ABs, since $\bm\chi_3$, $\bm\chi_4$ become almost real ($\bm\chi_{3i}\rightarrow0$, $\bm\chi_{4i}\rightarrow0$) when $\sigma=-1$, see Eqs. (\ref{eqchi-SR0}). As a result, the nonlinear superposition of two quasi-ABs with eigenvalues $\bm\chi_1$, $\bm\chi_2$ can produce a vector SR breather only in the defocusing regime.

However, this should be further substantiated by considering the exact relation between the MI and the SR breathers.
Namely, we proceed to calculate the absolute difference of the group velocities of the paired quasi-ABs: $\Delta V_g(\bm\chi_{1};\bm\chi_{2})=\bm|V_g(\bm\chi_1)-V_g(\bm\chi_2)\bm|$.
From Eq. (\ref{vg0}), we have
\begin{equation}
\Delta V_g(\bm\chi_{1};\bm\chi_{2})=\bm\left|\frac{\omega}{2\bm{\gamma}}(\bm\chi_{2i}-\bm\chi_{1i}) + \bm\chi_{2r}-\bm\chi_{1r}\bm\right|.\label{eqvg1}
\end{equation}
By inserting (\ref{eqchi-SR0}) into (\ref{eqvg1}) and omitting the higher-order terms, we have
\begin{equation}
\Delta V_g(\bm\chi_{1};\bm\chi_{2})=\bm\left| \frac{\omega}{\bm\gamma}\bm\chi^{(-)}_{_{AB}}\bm\right| .\label{eqvg2}
\end{equation}
Comparing the expressions for $\mathcal{G}$ and $\Delta V_g$, we have
\begin{equation}
\mathcal{G}=\bm\gamma\Delta V_g(\bm\chi_{1};\bm\chi_{2}).\label{eqlyy1}
\end{equation}
Equation (\ref{eqlyy1}) describes the explicit relation between the MI and the SR breathers that we obtained for Mahakov equations in the defocusing case.
This is the solid evidence of the existence of the vector SR breathers that has been  hidden previously due to the complexity of the Manakov equations.

The higher-order exact SR-breather solutions ($N$th-order solutions where $N$ is an even number) can be constructed by performing the next iterations of the Darboux transformation.
The details are shown in Supplemental Material.
Such iteration directly leads to the nonlinear superposition of the fundamental quasi-ABs, where each breather is associated with an individual eigenvalue and free parameters $(x_{0n},t_{0n}, \theta_n)$, $n=1,...,N$. If $\theta_n$ is fixed, the spatiotemporal distribution of such SR breather strongly depends on the relative separations in both $x$ and $t$, i.e., $\delta x=\{x_{01}, ..., x_{0N}\}$, and $\delta t=\{t_{01}, ..., t_{0N}\}$.

Alternatively,  the vector SR breathers can be observed in numerical simulations.
They can be excited using approximate initial conditions in the form of a localised modulation of a pane wave.
These can be called \emph{non-ideal} initial conditions, which do not necessarily follow from the exact solutions. They are more general localised perturbations of the vector plane-wave field:
\begin{equation}\label{eqic}
\bm\psi^{(j)}=\bm\psi_{0}^{(j)}\left[1+\epsilon L_p(x/x_{_W})\cos(\bm\omega x) \exp[{i\phi^{(j)}}]\right],
\end{equation}
where the localised function $L_p(x/x_{_W})$ is either the sech-function   $L_p=\textmd{sech}(x/x_{_W})$ or a Gaussian function $L_p=\exp{(-x^2/x_{_W}^2)}$ with $x_{_W}$ in each case denoting the width of the localisation. The width $x_{_W}$ must be comparable to that of our exact solutions, i.e., $1/(2\bm\gamma)$. The real parameters $\epsilon(\ll1)$ and $\bm\omega$ are the modulation amplitude and the frequency, respectively. The real constants $\phi^{(j)}$ denote the arbitrary phases.
Our numerical simulations show that once $x_{_W}$ and $\bm\omega$ are fixed by the exact solutions, the real phases $\phi^{(j)}$ are not  essential for the vector SR-breather excitation. We have found that the changes of $\phi^{(j)}$ only shift the position of the fundamental quasi-ABs [see Supplemental Material for details]. Thus we let $\phi^{(j)}=0$ in our simulations.

Figure \ref{f-defocusing} displays the evolution of the two components of the amplitude profiles of the vector field obtained from the numerical simulations and from the exact SR-breather solutions in the defocusing regime.
All breathers are dark structures due to the defocusing nonlinearity.
Remarkably, such solutions are absent in the case of the scalar NLSE.
Figure \ref{f-defocusing}(a) shows the two components of the SR-breather formation in the numerical simulations developed from the initial condition (\ref{eqic}). As can be seen from this figure, the initial perturbation is unstable relative to the modulation and evolves into two dark quasi-ABs with opposite velocities. The velocities are the same as in the exact solutions, i.e. $V_g(\bm\chi_1)$ and $V_g(\bm\chi_2)$. For comparison, Figure \ref{f-defocusing}(b) shows the exact SR-breather solution $\bm\psi^{(j)}(\bm\chi_{1};\bm\chi_{2})$.
Despite using the approximate initial conditions, the numerical simulations and the exact solution are in good agreement.
Once the parameters ($\bm\omega$, $x_{_W}$) are taken from the exact solutions,
the simple input (\ref{eqic}) can generate the SR breathers with good accuracy. Requirements to the accuracy of the initial conditions are sufficiently relaxed which means that the SR breathers are robust and can be easily excited. Moreover, as we demonstrate below, such a simple input can produce even more unexpected results in the focusing case.

\textit{Vector SR breathers and their excitation in the focusing regime}{.---} Next, we consider the SR breather formation in the focusing regime ($\sigma=1$) of the Manakov system.
For the focusing case, two different regimes can be distinguished: $\bm{\nu}|_{\bm{\gamma}=0}\geq0$ and $\bm{\nu}|_{\bm{\gamma}=0}<0$.
Below we discuss this point in more detail.

When $\bm{\nu}|_{\bm{\gamma}=0}\geq0$, implying that $\beta^2\leq a^4/(4a^2-\bm\omega^2)$, we have $\bm\chi^{(-)}_{_{AB}}\neq\bm\chi^{(+)}_{_{AB}}$.
However, we know that $\bm\chi^{(+)}_{_{AB}}=0$ when $\beta=0$. This fact is inconsistent with the decoupling case of the Manakov system ($\beta=0$) when the vector breather is reduced to the scalar one with a complex eigenvalue.
Thus, in the case $\bm{\nu}|_{\bm{\gamma}=0}\geq0$,
only $\bm{\chi}_{1}$ and $\bm{\chi}_{2}$ are valid eigenvalues for the quasi-ABs.
Then, the interaction between the two quasi-ABs, [$\bm\psi^{(j)}(\bm\chi_1;\bm\chi_2$)], leads to the formation of the vector SR-breather. This happens in the focusing regime
 when $\bm{\nu}|_{\bm{\gamma}=0}\geq0$  just as in the defocusing case [see examples in Supplemental Material].

On the contrary, when $\bm{\nu}|_{\bm{\gamma}=0}<0$, implying that $\beta^2>a^4/(4a^2-\bm\omega^2)$, we have $\bm\chi^{(-)}_{_{AB}}=-\bm\chi^{(+)}_{_{AB}}$, and $\bm{\rho}^{(-)}=\bm{\rho}^{(+)}$.
In this case, all four eigenvalues (\ref{eqchi-SR0}) are valid for the formation of the quasi-ABs.
However, not every nonlinear superposition leads to the formation of the
quasi-ABs.
In particular, from Eq. (\ref{eqchi-SR0}) we have
\begin{equation}\label{eq-vg-f}
\Delta V_g(\bm\chi_{1};\bm\chi_{3})=\Delta V_g(\bm\chi_{2};\bm\chi_{4})=\bm\left|\frac{\omega}{\bm\gamma}\bm\chi^{(-)}_{_{AB}}\bm\right|.
\end{equation}
Then, in contrast to all cases considered above, there are two SR breathers sharing the same growth rate for any given localised periodic modulation with the width $1/\bm\gamma$ and the frequency $\omega$. The growth rate is:
\begin{equation}
\mathcal{G}=\bm\gamma\Delta V_g(\bm\chi_{1};\bm\chi_{3})=\bm\gamma\Delta V_g(\bm\chi_{2};\bm\chi_{4}).\label{lyy2}
\end{equation}
Such variety of the combinations unveils the richness of the dynamics of the vector SR breathers.

\begin{figure}[htb]
\centering
\includegraphics[width=0.5\textwidth]{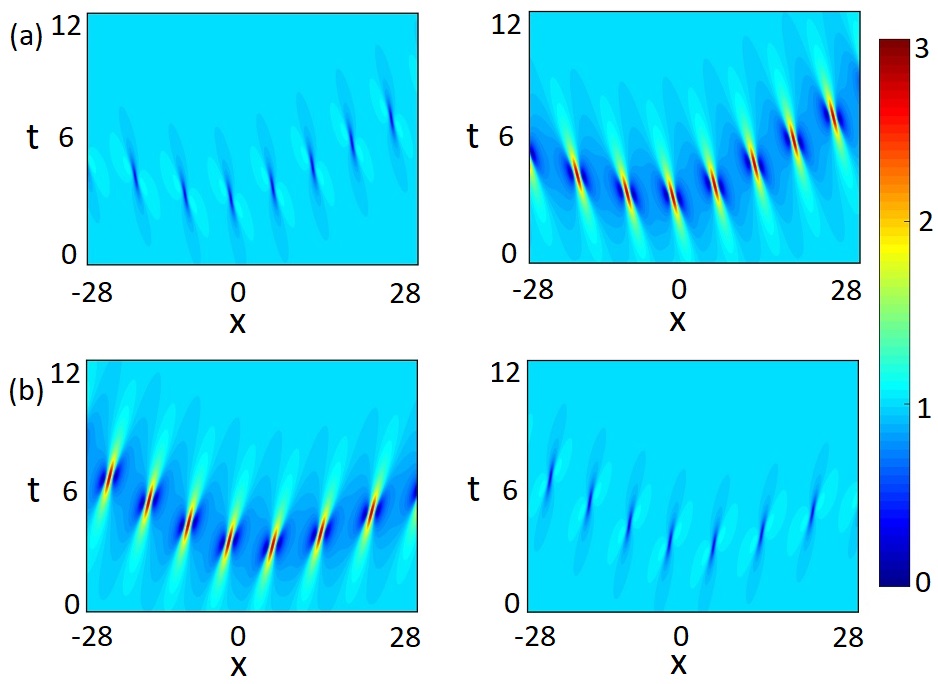}
\caption{Amplitude evolution of the two exact solutions (a) $|\bm\psi^{(j)}(\bm\chi_{1};\bm\chi_{3})|$ and (b) $|\psi^{(j)}(\bm\chi_{2};\bm\chi_{4})|$ of the vector SR breathers
in the focusing regime ($\sigma=1$). Parameters are: $a_1=a_2=1$, $\beta=1$, $\bm\gamma=0.1$, $\bm\omega=0.8$, $\theta_n=0$, $\delta x=\{-5.9118,-2.5549\}$, $\delta t=\{-1.2720,-0.7061\}$.}\label{f-focusing}
\end{figure}

Figures \ref{f-focusing}(a) and \ref{f-focusing}(b) show the amplitude evolution of the two components of these two SR breathers, $|\psi^{(j)}(\bm\chi_{1};\bm\chi_{3})|$ and $|\psi^{(j)}(\bm\chi_{2};\bm\chi_{4})|$ respectively. For a given vector plane-wave field defined by the amplitudes $a_1$ and $a_2$, the SR breather $|\psi^{(j)}(\bm\chi_{1};\bm\chi_{3})|$ exhibits the quasi-AB with dark envelope distribution in $\psi^{(1)}$ and bright envelope distribution in $\psi^{(2)}$ components. On the contrary,
 the SR breather $|\psi^{(j)}(\bm\chi_{2};\bm\chi_{4})|$ shows the bright envelope distribution in $\psi^{(1)}$ and the dark envelope distribution in $\psi^{(2)}$ components.
In each case, the two quasi-ABs are asymmetric with respect to the reversing the axis $x\rightarrow -x$. This is due to the unequal group velocities $V_g(\bm\chi_{1})\neq -V_g(\bm\chi_{3})$ and $V_g(\bm\chi_{2})\neq -V_g(\bm\chi_{4})$. However, the absolute values of the differences of the group velocities in the two cases are the same as in Eq. (\ref{eq-vg-f}).
Nevertheless, the two types of SR breathers require different initial conditions at $t=0$. Each type of the SR breathers can be excited in numerical simulations that start from the individual initial conditions $\bm\psi^{(j)}(0,x)$.

\begin{figure}[htb]
\centering
\includegraphics[width=0.5\textwidth]{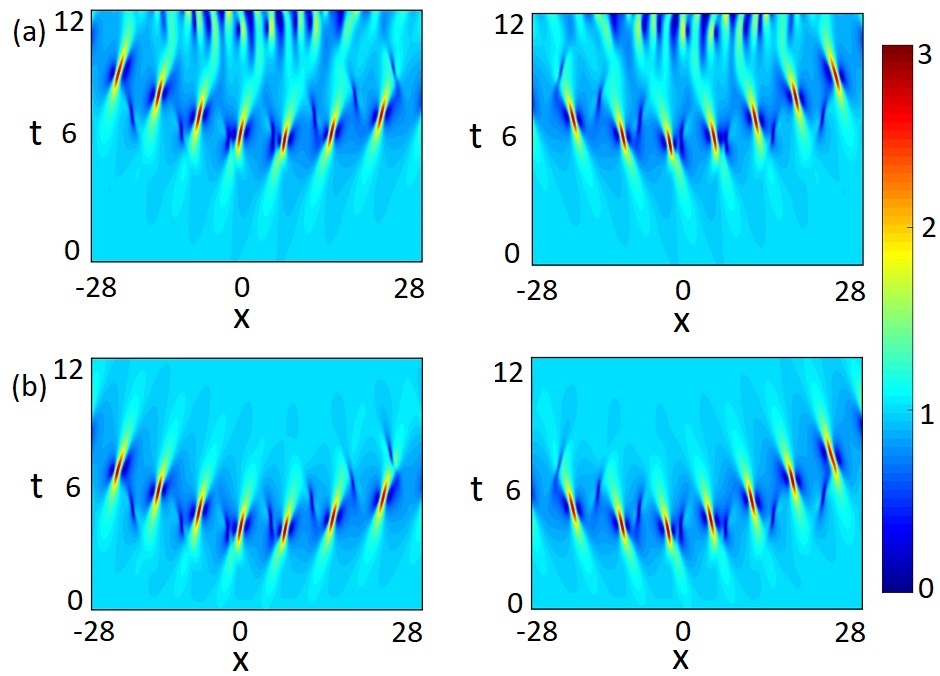}
\caption{Evolution of the amplitude profiles $|\bm\psi^{(j)}|$ of the vector SR breathers
in the focusing regime ($\sigma=1$). (a) Numerical simulations started with the initial conditions (\ref{eqic}). (b) Exact solution $|\psi^{(j)}(\bm\chi_{1};\bm\chi_{3};\bm\chi_{2};\bm\chi_{4})|$.
Parameters are: $a_1=a_2=1$, $\beta=1$, $\bm\gamma=0.1$, $\bm\omega=0.8$, $\delta x_{(1,3,2,4)}=\{-6.9118, -11.2814,-6.2324,-5.0731\}$, $\delta t_{(1,3,2,4)}=\{-0.5220, 1.2178, 1.2620, -0.4747\}$, $\theta_n=0$, and $x_{_W}=10$, $\phi^{(j)}=0$, $\epsilon=0.01$.}\label{f-focusing-2}
\end{figure}

Now, the next question is: can these two SR breathers coexist on the same background? Our studies show that the answer to this question is affirmative. Indeed, the fourth-order iteration of the Darboux transformation provides the corresponding exact solution. These solutions vary and the specific SR structure depends on the free parameters $(x_{0n},t_{0n})$ for each breather. These complex structures can be also excited in numerical simulations. The remarkable finding here is that such complex structures can be excited using the same simple initial conditions (\ref{eqic}) that generate lower order SR breathers.

Figures \ref{f-focusing-2}(a) and \ref{f-focusing-2}(b) show the amplitude evolution of higher-order SR-breathers that are obtained in numerical simulations and from the exact solution respectively.
The numerical simulations of such complex SR breathers shown in Fig. \ref{f-focusing-2}(a) start from the localised small amplitude MI and evolve into the nonlinear superposition of four breathers rather than two. This pattern can also be considered as a superposition of the two types of SR breathers shown in Fig. \ref{f-focusing}.
Our exact fourth-order solution $|\psi^{(j)}(\bm\chi_{1};\bm\chi_{3};\bm\chi_{2};\bm\chi_{4})|$ with carefully chosen parameters confirms the results of our numerical simulations.
Indeed, the main features of the patterns in Figs. \ref{f-focusing-2}(a) and \ref{f-focusing-2}(b) are similar. The differences of the evolution in the numerical simulations after the excitation of the higher-order SR breather are caused by the inaccuracy of the initial conditions. We should remember though that the excitation of the higher-order structures using the simple initial conditions is possible only when the two lowest-order SR breathers share the same MI growth rate.


\textit{Conclusions}{.---} Here, we presented the theory of the SR breathers for Manakov equations. We considered the cases of both the focusing and the defocusing regimes.
These studies are based on the eigenvalue analysis and the expressions for the MI growth rates of the SR breathers. We confirmed the results of the analytical investigation
by numerical simulations. In particular, we have shown that a simple localised initial modulation may lead to the excitation of both a single SR breather as well as the second-order SR breather. The latter involves four fundamental vector breathers.

Our theory reveals the properties of the SR breathers that can be useful for
understanding their physics in systems described by the Manakov equations.
This, in turn, can be useful in practical applications and for the interpretation of experimental results.

The work of Liu is supported by the NSFC (Grants No. 12247103, and No. 12175178),
the Natural Science basic Research Program of Shaanxi Province (Grant No. 2022KJXX-71), and Shaanxi Fundamental
Science Research Project for Mathematics and Physics (Grant No. 22JSY016).
The work of Akhmediev is supported by the Qatar National Research Fund (grant NPRP13S-0121-200126).

\end{CJK}

\end{document}